\documentclass{ws-ijmpcs}

\usepackage[english]{babel}
\usepackage{enumerate}
\usepackage{amsmath,amssymb}
\usepackage{letltxmacro}			
\usepackage{xkeyval}				
\usepackage{xargs}					
\usepackage{xifthen}				
\usepackage{slashed}
\usepackage[normalem]{ulem}
\usepackage{bm}					
\usepackage{dsfont}
\usepackage[super,sort&compress]{natbib}
\usepackage[cmyk,usenames,dvipsnames,svgnames,table,x11names]{xcolor}
\usepackage{tikz}
\usepackage{tikz-3dplot}
\usetikzlibrary{calc,matrix, shapes,arrows,decorations.pathmorphing,decorations.markings,decorations.pathreplacing,trees}
\usepackage[pdftex, unicode=true, pdfnewwindow=true, colorlinks=true, linkcolor=black, urlcolor=black, citecolor=black, hyperindex, pdfauthor={Frederik F. Van der Veken}, pdftitle={Working with Piecewise Linear Wilson Lines}, bookmarksopen=true, bookmarksopenlevel=0]{hyperref}

\makeatletter
	\let\oldr@@t\r@@t
	\def\r@@t#1#2{%
	\setbox0=\hbox{$\oldr@@t#1{#2\,}$}\dimen0=\ht0
	\advance\dimen0-0.2\ht0
	\setbox2=\hbox{\vrule height\ht0 depth -\dimen0}%
	{\box0\lower0.4pt\box2}}
	\LetLtxMacro{\oldsqrt}{\sqrt}
	\renewcommand*{\sqrt}[2][\ ]{\oldsqrt[#1]{#2}}
\makeatother

\newcommandx*{\1}[2][1,2]{\ensuremath{
		\ifthenelse{\isempty{#1} \AND \isempty{#1}}{\mathds{1}}{
			\ifthenelse{\isempty{#2}}
				{\mathds{1}_{#1}}
				{\mathds{1}_{{#1}\times {#2}}}
}
}}

\newcommand{\ep}{\, .}
\newcommand{\ec}{\, ,}

\renewcommand*{\bar}{\overline}

\newcommand{\Not}{\stackrel{\text{\tiny N}}{=}}							
\renewcommand{\exp}[1]{\ensuremath{\text{e}^{#1}}}

\renewcommand*{\i}{\text{i}\hspace*{1pt}}

\newlength{\nplength}
\newcommandx{\negphantom}[1]{\settowidth{\nplength}{#1}\hspace{-\nplength}}

\renewcommandx*{\vec}[3][2, 3]{{\ensuremath{\mathbf{\bm{#1}}_\mathrm{#2}^{\mathrm{#3}}}}}
\newcommandx*{\Trace}[1][1]{\ensuremath{
		\:\textrm{Tr}\ifthenelse{\isempty{#1}}{\,}{\!\left( #1 \right)}
}}
\newcommandx*{\trace}[1][1]{\ensuremath{
		\:\textrm{tr}\ifthenelse{\isempty{#1}}{\,}{\!\left( #1 \right)}
}}

\newcommandx*{\diracdelta}[2][1]{\ensuremath{
		\ifthenelse{\isempty{#1}}
		{\,\delta\!\left({#2}\right)}
		{\,\delta^{\IfInteger{#1}{({#1})}{{#1}}}\!\left({#2}\right)}
}}

\newcommandx*{\Int}[4][1, 2, 4={0pt}, usedefault, addprefix=\global]{
	\ensuremath{\int\limits_{\:\!{#1}}^{\:\!{#2}}{\!\!}\hspace{#4} {\protect#3}\,\,}
}

\newcommandx*{\Dif}[2][1]{\ensuremath{{\textrm{d}}^{{#1}} {#2}\,}}
\newcommandx*{\dif}[2][1]{\ensuremath{\partial^{{#1}} {#2}\,}}
\newcommandx*{\MDif}[2][1]{
	\ensuremath{\frac{{\textrm{d}}^{{#1}} {#2}\,}{
		\ifthenelse{\isempty{#1}}{2\pi} { \left(2\pi\right)^{{#1}} }
	}}
}
\newcommandx*{\DDif}[2][1]{\ensuremath{{\mathcal{D}}^{{#1}} {#2}\,}}
\newcommandx*{\Diff}[8][1, 4, 5, 6, 7, 8]{\ensuremath{
		\ifthenelse{\isempty{#4} \AND \isempty{#5}}
			{\frac{\Dif[#1]{#2}}{{\Dif{#3}}^{{#1}}}}
			{\frac{\Dif[#1]{#2}}{
				\ifthenelse{\isempty{#4}} {\Dif{#3}} {\left(\Dif{#3}\right)^{#4}}
				\ifthenelse{\isempty{#5}} {} { \ifthenelse{\isempty{#6}} {\Dif{#5}} {\left(\Dif{#5}\right)^{#6}}}
				\ifthenelse{\isempty{#7}} {} { \ifthenelse{\isempty{#8}}{\Dif{#7}}{\left(\Dif{#7}\right)^{#8}}}
}}}}
\newcommandx*{\diff}[8][1, 4, 5, 6, 7, 8]{\ensuremath{
		\ifthenelse{\isempty{#4} \AND \isempty{#5}}
			{\frac{\dif[#1]{#2}}{{\dif{#3}}^{{#1}}}}
			{\frac{\dif[#1]{#2}}{
				\ifthenelse{\isempty{#4}} {\dif{#3}} {\left(\dif{#3}\right)^{#4}}
				\ifthenelse{\isempty{#5}} {} { \ifthenelse{\isempty{#6}} {\dif{#5}} {\left(\dif{#5}\right)^{#6}}}
				\ifthenelse{\isempty{#7}} {} { \ifthenelse{\isempty{#8}}{\dif{#7}}{\left(\dif{#7}\right)^{#8}}}
}}}}

\newcommandx*{\braket}[3][1,2,3]{
	\ifthenelse{\isempty{#2} \AND \isempty{#3}}
	{\ifthenelse{\isempty{#1}}  {\opm{!! empty braket used !!}}  {\ensuremath{\left< {#1} \right>} } }
	{\ifthenelse{\isempty{#3}}
			{\ensuremath{\left< {#1} \vphantom{{#2}}  \:\! \right| \! \! \! \; \left. {#2} \vphantom{{#1}} \right>} }
			{\ensuremath{\left< {#1} \vphantom{{#3}} \right| #2 \left| {#3} \vphantom{{#1}} \right>} }
}}

\newcommandx*{\wilson}[3][3=]{\ensuremath{
	\,\ifthenelse{\isempty{#1} \AND \isempty{#2}}	{\mathcal{U}^{#3}}	{\mathcal{U}_{(#1 \, ; \,  #2)}^{#3}}
}}

\newcommand{\wilsonup}{
	\def\myscale{0.6}
	\begin{tikzpicture}[scale=0.6, baseline= {($(current bounding box.base)-(0,1.7pt)$)}]
		\draw[wilson,-implies] (0,0) -- (1.5,0);
		\filldraw[wilsontext] (0,0) circle(0.125);
	\end{tikzpicture}
}
\newcommand{\wilsondown}{
	\def\myscale{0.6}
	\begin{tikzpicture}[scale=0.6, baseline= {($(current bounding box.base)-(0,1.7pt)$)}]
		\draw[wilson,implies-] (0,0) -- (1.5,0);
		\filldraw[wilsontext] (1.5,0) circle(0.125);
	\end{tikzpicture}
}
\newcommand{\wilsonupreversed}{
	\def\myscale{0.6}
	\begin{tikzpicture}[scale=0.6, baseline= {($(current bounding box.base)-(0,1.7pt)$)}]
		\draw[wilson] (0,0) -- (1.5,0);
		\draw[wilson,-implies] (1.56,0) -- (1.4,0);
		\filldraw[wilsontext] (0,0) circle(0.125);
	\end{tikzpicture}
}
\newcommand{\wilsondownreversed}{
	\def\myscale{0.6}
	\begin{tikzpicture}[scale=0.6, baseline= {($(current bounding box.base)-(0,1.7pt)$)}]
		\draw[wilson] (0,0) -- (1.5,0);
		\draw[wilson,-implies] (-0.06,0) -- (0.1,0);
		\filldraw[wilsontext] (1.5,0) circle(0.125);
	\end{tikzpicture}
}

\newenvironment{subalign}[1][]{
	\subequations
	\ifthenelse{\isempty{#1}}{}{\label{#1}}
	\align
}{
	\endalign
	\endsubequations
}

\definecolor{geel}{cmyk}{0.,0.075,1.,0.2}
\definecolor{blauw}{cmyk}{1.,0.55,0.05,0.05}
\definecolor{groen}{cmyk}{0.6,0.15,1.,0.1}
\definecolor{rood}{cmyk}{0.1,1.,0.1,0.3}
\newcommand{\photoncolour}{geel}
\newcommand{\gluoncolour}{groen}
\newcommand{\fermioncolour}{geel}
\newcommand{\darkfermioncolour}{geel!90!blauw!80!black}
\newcommand{\wilsoncolour}{blauw}
\newcommand{\eikonalcolour}{rood}
\newcommand{\bloboutercolour}{gray!80!black}
\newcommand{\blobinnercolour}{gray!20}
\newcommand{\pdfoutercolour}{geel!80!groen!80!blauw}
\newcommand{\pdfinnercolour}{geel!60!groen!80!blauw!20!white}

\pgfdeclaredecoration{complete sines}{initial}{				
	\state{initial}[
		width=+0pt,
		next state=sine,
		persistent precomputation={
			\pgfmathsetmacro\matchinglength{
				\pgfdecoratedinputsegmentlength / int(\pgfdecoratedinputsegmentlength/\pgfdecorationsegmentlength)
			}
			\setlength{\pgfdecorationsegmentlength}{\matchinglength pt}
        		}
	] {}
	\state{sine}[width=\pgfdecorationsegmentlength]{
		\pgfpathsine{\pgfpoint{0.25\pgfdecorationsegmentlength}{0.5\pgfdecorationsegmentamplitude}}
		\pgfpathcosine{\pgfpoint{0.25\pgfdecorationsegmentlength}{-0.5\pgfdecorationsegmentamplitude}}
		\pgfpathsine{\pgfpoint{0.25\pgfdecorationsegmentlength}{-0.5\pgfdecorationsegmentamplitude}}
		\pgfpathcosine{\pgfpoint{0.25\pgfdecorationsegmentlength}{0.5\pgfdecorationsegmentamplitude}}
	}
	\state{final}{}
}
\pgfdeclaredecoration{integralshape}{integralshape}{		
	\state{integralshape}[width=+\pgfdecoratedremainingdistance,next state=final] {
		\pgfpathmoveto{\pgfpoint{0}{0.042*\pgfdecoratedpathlength*\pgfdecorationsegmentamplitude}}
		\pgfpathcurveto
			{\pgfpoint{-0.01*\pgfdecoratedpathlength*\pgfdecorationsegmentamplitude}{0.02*\pgfdecoratedpathlength*\pgfdecorationsegmentamplitude}}
			{\pgfpoint{\pgfdecoratedpathlength*(0.5-0.18*\pgfdecorationsegmentamplitude)}{0.18*tan(\pgfdecorationsegmentangle)*\pgfdecorationsegmentamplitude*\pgfdecoratedpathlength}}
			{\pgfpoint{\pgfdecoratedpathlength*0.5}{0}}
		\pgfpathcurveto
			{\pgfpoint{\pgfdecoratedpathlength*(0.5+0.18*\pgfdecorationsegmentamplitude)}{-0.18*tan(\pgfdecorationsegmentangle)*\pgfdecorationsegmentamplitude*\pgfdecoratedpathlength}}
			{\pgfpoint{\pgfdecoratedpathlength*(1+0.01*\pgfdecorationsegmentamplitude)}{-0.02*\pgfdecoratedpathlength*\pgfdecorationsegmentamplitude}}
			{\pgfpoint{\pgfdecoratedpathlength}{-0.042*\pgfdecoratedpathlength*\pgfdecorationsegmentamplitude}}
	}
	\state{final}{}
}

\def\myscale{1}
\tikzset{
photon/.style={decorate, decoration={snake,amplitude=\myscale*3pt, segment length=\myscale*7pt}, draw=\photoncolour, line width=\myscale*1.4pt},
gluon/.style={decorate, draw=\gluoncolour, decoration={coil,amplitude=\myscale*3pt, segment length=\myscale*4pt},  line width=\myscale*1.2pt},
quark/.style={draw=\fermioncolour, postaction={decorate},
	decoration={markings,mark=at position .5*\pgfdecoratedpathlength+sqrt(\myscale)*5.5pt with {\arrow[\fermioncolour]{latex}}},  line width=\myscale*1.6pt},		
quarknoarrow/.style={draw=\fermioncolour, line width=\myscale*1.6pt},
darkquark/.style={draw=\darkfermioncolour, postaction={decorate},
	decoration={markings,mark=at position .5*\pgfdecoratedpathlength+sqrt(\myscale)*5.5pt with {\arrow[\darkfermioncolour]{latex}}},  line width=\myscale*1.6pt},
eikonal/.style={double, double distance=\myscale*2.5pt, draw=\eikonalcolour, postaction={decorate},
	decoration={markings,mark=at position .5*\pgfdecoratedpathlength+sqrt(\myscale)*7.5pt with {\arrow[\eikonalcolour]{angle 60}}}, line width=\myscale*1.2pt},
wilson/.style={double, double distance=\myscale*2.5pt, line width=\myscale*1.2pt, draw=\wilsoncolour},
blob/.style={draw=\bloboutercolour, fill=\blobinnercolour, line width=\myscale*1.2pt},
pdf/.style={draw=\pdfoutercolour, fill=\pdfinnercolour, line width=\myscale*1.2pt},
photontext/.style={\photoncolour!80!black},
gluontext/.style={\gluoncolour!80!black},
quarktext/.style={\fermioncolour!80!black},
wilsontext/.style={\wilsoncolour!80!black},
eikonaltext/.style={\eikonalcolour!80!black},
blobtext/.style={\bloboutercolour!80!black},
pdftext/.style={\pdfoutercolour!80!black},
accolade/.style={gray,decorate, decoration={brace,amplitude=\myscale*5pt}, line width=\myscale*1.2pt},
hide/.style={ultra thick, white},
finalstatecut/.style={decorate, decoration={integralshape,amplitude=(1/\myscale)*1.5pt,angle=2}, line width=\myscale*1.2pt},
wilsonarrow/.style={postaction={decorate}, decoration={markings,mark=at position .75 with {\arrow[\wilsoncolour]{angle 60}}}},
wilsonarrowreversed/.style={postaction={decorate}, decoration={markings,mark=at position .3 with {\arrowreversed[\wilsoncolour]{angle 60}}}},
wilsonarrow2/.style={postaction={decorate}, decoration={markings,mark=at position 1. with {\arrow[\wilsoncolour]{angle 60}}}},
wilsonarrowreversed2/.style={postaction={decorate}, decoration={markings,mark=at position 0 with {\arrowreversed[\wilsoncolour]{angle 60}}}}
}
\newenvironment{tikzfigure}[2]{
	\def\myscale{#1}
	\begin{tikzpicture}[baseline= {($(current bounding box.base)-(0pt,#2)$)},scale=\myscale]
}
{
	\end{tikzpicture}
}

\newcommand{\Eg}{E.g.\ }
\newcommand{\eg}{e.g.\ }

\begin{document}

\markboth{Frederik~F.~Van~der~Veken}{Working With Wilson Lines}

%
\catchline{}{}{}{}{}
%

\title{Working With Wilson Lines}

\author{Frederik~F.~Van~der~Veken}

\address{Department of Physics, University of Antwerp, Groenenborgerlaan 171, 2020 Antwerp, Belgium\\
frederikvanderveken@gmail.com}

\maketitle


\begin{abstract}
We present an algorithm to express Wilson lines that are defined on piecewise linear paths in function of their individual segments, reducing the number of diagrams needed to be calculated. The important step lies in the observation that different linear path topologies can be related to each other using their color structure. This framework allows one to easily switch results between different Wilson line topologies, which is helpful when testing different structures against each other.
\end{abstract}

\ccode{PACS numbers: 11.15.Tk, 12.38.Aw, 12.38.Lg.}

\section{Introduction}
Wilson lines, path-ordered exponentials of the gauge field, are generally defined along a path $\mathcal{C}$ as:
\begin{subalign}
  \mathcal{U} &= \nonumber
    \mathcal{P}\,\exp{\i g \int_\mathcal{C} \Dif{z^\mu} A_\mu (z)}
    \ec \\ \label{eq: Wilsondef}
  &=
    \sum_{n=0}^\infty
    \left(\i g\right)^n
    \Int{\MDif[\omega]{k_1}\cdots \MDif[\omega]{k_n}}
    A_{\mu_n}(-k_n)\cdots A_{\mu_1}(-k_1) \;
    I_n
    \ec \\
  I_n &=
    \frac{1}{n!}\; 
    \mathcal{P} \!\!\int\! \Dif{\lambda_1}\cdots \Dif{\lambda_n}
    \left(z_1^{\mu_1}\right)' \cdots \left(z_n^{\mu_n}\right)'
    \exp{\i\prod\limits^n k_i\cdot z_i}\ec
\end{subalign}
where the symbol $\mathcal{P}$ denotes path ordering and $\lambda$ is a parameterization of the path. A Wilson line's gauge transform only depends on the endpoints, which is a property that can be used to give gauge invariant definitions for bilocal operators.\cite{Collins:1981uk,Collins:1982wa,Collins:1984kg,Belitsky:2002sm,Boer:2003cm,Ji:2004xq,Bacchetta:2004jz,Belitsky:2005qn,Hautmann:2007uw,Cherednikov:2009wk,Cherednikov:2011ku,Collins:2011zzd,Boer:2011fh,MertAybat:2011wy,Echevarria:2012js,Bacchetta:2012qz,Collins:2013aa} The applicability of Wilson lines is very broad. \Eg in quantum chromodynamics (QCD) they are used---among others---in the study of jet quenching\cite{Cherednikov:2013pba} and as basic ingredients for a framework where QCD is fully recast in loop space based on a geometric evolution.\cite{Mertens:2014hma,Cherednikov:2012qq,VanderVeken:2014lka,Mertens:2014lla,Mertens:2014mia}

\section{Linear Path Segments}
\begin{figure*}[t!]
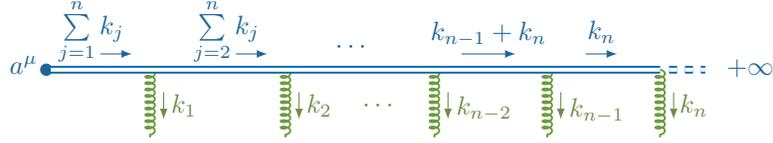

\centering
\begin{tikzfigure}{0.6}{0}
  \node[wilsontext] at (17.3,0.03) {$+\infty$};
  \filldraw[wilsontext] (1.7,0) circle(0.125);
  \node[wilsontext] at (1.2,0.1) {$a^\mu$};
  \draw[gluon] (4,0) -- (4,-1.5);
  \draw[-latex, gluontext] (4.3,-0.5) -- (4.3, -1.1);
  \node[gluontext] at (4.75,-0.8) {$k_1$};
  \draw[-latex,wilsontext] (2.9,0.35) -- (3.6,0.35);
  \node[wilsontext] at (2.7,0.85) {$\sum\limits_{j=1}^{n} k_j $};
  \begin{scope}[shift={(3,0)}]
    \draw[gluon] (4,0) -- (4,-1.5);
    \draw[-latex, gluontext] (4.3,-0.5) -- (4.3, -1.1);
    \node[gluontext] at (4.75,-0.8) {$k_2$};
    \draw[-latex,wilsontext] (2.9,0.35) -- (3.6,0.35);
    \node[wilsontext] at (2.7,0.85) {$\sum\limits_{j=2}^{n} k_j $};
  \end{scope}
  \node[wilsontext] at (8.5,0.5) {$\cdots$};
  \node[gluontext] at (9.1,-0.8) {$\cdots$};
  \begin{scope}[shift={(6.3,0)}]
    \draw[gluon] (4,0) -- (4,-1.5);
    \draw[-latex, gluontext] (4.3,-0.5) -- (4.3, -1.1);
    \node[gluontext] at (5.1,-0.85) {$k_{n-2}$};
  \end{scope}
  \begin{scope}[shift={(8.8,0)}]
    \draw[gluon] (4,0) -- (4,-1.5);
    \draw[-latex, gluontext] (4.3,-0.5) -- (4.3, -1.1);
    \node[gluontext] at (5.1,-0.9) {$k_{n-1}$};
    \draw[-latex,wilsontext] (2.1,0.35) -- (3.3,0.35);
    \node[wilsontext] at (2.7,0.8) {$ k_{n-1}+k_n $};
  \end{scope}
  \begin{scope}[shift={(11.3,0)}]
    \draw[gluon] (4,0) -- (4,-1.5);
    \draw[-latex, gluontext] (4.3,-0.5) -- (4.3, -1.1);
    \node[gluontext] at (4.75,-0.8) {$k_n$};
    \draw[-latex,wilsontext] (2.35,0.35) -- (3.05,0.35);
    \node[wilsontext] at (2.7,0.85) {$k_n $};
  \end{scope}
  \draw[wilson] (1.8,0) -- (15.3,0);
  \draw[wilson,dashed] (15.5,0) -- (16.3,0);
\end{tikzfigure}
\caption{$n$-gluon radiation for a Wilson line going from $a^\mu$ to $+\infty$.}
\label{fig: lowerbound}
\end{figure*}

Before we go to more complex calculations, we review the calculation of one linear segment. There are four possible linear path structures: it can be a finite path, a semi-infinite line, or a fully infinite line. In this paper we don't treat the last case.
We start  with a semi-infinite line from a point $a^\mu$ to $+\infty$ along a direction $\hat{n}^\mu$:
\begin{equation}
  z^\mu = a^\mu + \lambda \, \hat{n}^\mu \quad \lambda = 0 \ldots \infty\ep
\end{equation}
It is then straightforward to calculate the segment integrals (see Fig.\@ \ref{fig: lowerbound}):
\begin{equation}
  I_n^\text{l.b.} =  \label{eq: lowerbound}
    \hat{n}^{\mu_1}\cdots \hat{n}^{\mu_n} \,
    \exp{\i a\cdot \sum\limits_j k_j} \,
    \prod\limits_{j=1}^n
    \frac{\i}{\hat{n}\!\cdot\! \sum\limits_{l=j}^n k_l + \i \eta}\ep
\end{equation}
%
Next we investigate a path that starts at $-\infty$ and goes up to a point $b_\mu$:
\begin{equation}
  z^\mu = b^\mu  + \hat{n}^\mu \, \lambda \qquad \lambda = -\infty \ldots 0\ep
\end{equation}
The resulting path integral is almost the same as before:
\begin{equation}\label{eq: upperbound}
  I_n^\text{u.b.} =
    \hat{n}^{\mu_1}\cdots \hat{n}^{\mu_n} \,
    \exp{\i b\cdot \sum\limits_j k_j} \,
    \prod\limits_{j=1}^n
    \frac{-\i}{\hat{n}\!\cdot\! \sum\limits_{l=1}^j k_l - \i \eta} \ec
\end{equation}
which differs from \eqref{eq: lowerbound} only in the accumulation of momenta in the denominators.

We now introduce a shorthand notation to denote the path structure for a Wilson line segment. We represent the two structures we already calculated as:
\begin{align}
  \wilson{+\infty}{a} &\quad \Not \quad \wilsonup \ec &
  \wilson{b}{-\infty} &\quad \Not \quad \wilsondownreversed \ep 
\end{align}
Path reversing equals changing the type and flipping $\hat{n}$.\cite{VanderVeken:2014kna,VanderVeken:2014eda} We represent this as:
\begin{subalign}
  \wilson{a}{+\infty} = \wilson{a}{-\infty}\Big|_{\hat{n}\,\rightarrow\,-\hat{n}} &\quad \Not \quad  \wilsondown \ec \\
  \wilson{-\infty}{b} = \wilson{+\infty}{b}\Big|_{\hat{n}\,\rightarrow\,-\hat{n}} &\quad \Not \quad \wilsonupreversed \ep 
\end{subalign}
A nice feature of this notation is that we get a symbolic ``mirror relation'':
\begin{align}
  \left( \wilsonup\right)^\dagger
  &= \,
  \wilsondown
  \ec
  &
  \left(\wilsondownreversed\right)^\dagger
  &= \,
  \wilsonupreversed
  \ep
\end{align}
It can be shown\cite{VanderVeken:2014kna,VanderVeken:2014eda} that a finite line equals a product of two semi-infinite lines:
\begin{equation}
  \wilson{b}{a} = \wilson{+\infty}{b}[\dagger]\wilson{+\infty}{a} = \wilson{b}{-\infty}\wilson{a}{-\infty}[\dagger],
\end{equation}
which can be illustrated schematically as:
\begin{equation}
  \begin{tikzfigure}{0.6}{0.08}
    \draw[wilson, wilsonarrow] (0,0) -- (2,0);
    \filldraw[wilsontext] (0,0) circle(0.125);
    \filldraw[wilsontext] (2,0) circle(0.125);
  \end{tikzfigure}
  \quad = \quad \wilsonup \,\otimes \, \wilsondown
  \quad = \quad \wilsonup \, \otimes \left(\wilsonup\right)^\dagger \ep
\end{equation}

\section{Relating  Different Path Topologies}
We can trivially relate all six path structures to each other, except two 
which are related by a sign difference and an interchange of momentum indices:
\begin{equation}\label{eq: BasicWilsons}
  \wilsondown \; = \; (-)^n \,
  \wilsonup
  \big|_{\left(k_1,\ldots,k_n\right) \rightarrow \left(k_n,\ldots,k_1\right)}\ep
\end{equation}
We can exploit this relation when connecting a Wilson line to a blob:
\begin{equation*}
  \begin{tikzfigure}{0.6}{0.5}
    \draw[gluon] (0.5,0) -- (0.5,-1.5);
    \draw[gluon] (1,0) -- (1,-1.5);
    \draw[gluon] (2.5,0) -- (2.5,-1.5);
    \node[gluontext] at (1.75,-0.45) {$\ldots$};
    \draw[wilson, -implies] (0,0) -- (3,0);
    \filldraw[wilsontext] (0,0) circle(0.125);
    \filldraw[blob] (1.5,-1.5) circle(1.5 and 0.75);
    \node[blobtext] at (1.5,-1.5) {$F$};
  \end{tikzfigure}
  \! = \;
    \left(\i g\right)^n t^{a_n}\!\cdots t^{a_1}\!
    \Int{\MDif[\omega]{k_1}\cdots \MDif[\omega]{k_n}}
    \; I_n^\text{l.b.} \;\,
    F_{\mu_1\cdots\mu_n}^{a_1\cdots a_n}{ (k_1,\ldots, k_n)}
    \ep
\end{equation*}
The blob can contain anything but Wilson lines, and gluon propagators are automatically absorbed into the blob. It is defined as the sum of all possible crossings, making it symmetric under the simultaneous interchange of Lorentz, color, and momentum indices. As the Lorentz structure of the Wilson line is symmetric, that of $F$ is automatically symmetric as well. This then induces a color-momentum relation: an interchange of momentum variables is equivalent to an interchange of the corresponding color indices. Hence the two structures in Eq.\ \eqref{eq: BasicWilsons} are related by:
\begin{equation}
  \begin{tikzfigure}{0.6}{0.5}
    \draw[gluon] (0.5,0) -- (0.5,-1.5);
    \draw[gluon] (1,0) -- (1,-1.5);
    \draw[gluon] (2.5,0) -- (2.5,-1.5);
    \node[gluontext] at (1.75,-0.45) {$\ldots$};
    \draw[wilson, implies-] (0,0) -- (3,0);
    \filldraw[wilsontext] (3,0) circle(0.125);
    \filldraw[blob] (1.5,-1.5) circle(1.5 and 0.75);
    \node[blobtext] at (1.5,-1.5) {$F$};
  \end{tikzfigure}
  = (-)^n 
  \begin{tikzfigure}{0.6}{0.5}
    \draw[gluon] (0.5,0) -- (0.5,-1.5);
    \draw[gluon] (1,0) -- (1,-1.5);
    \draw[gluon] (2.5,0) -- (2.5,-1.5);
    \node[gluontext] at (1.75,-0.45) {$\ldots$};
    \draw[wilson, -implies] (0,0) -- (3,0);
    \filldraw[wilsontext] (0,0) circle(0.125);
    \filldraw[blob] (1.5,-1.5) circle(1.5 and 0.75);
    \node[blobtext] at (1.5,-1.5) {$F$};
  \end{tikzfigure}
  \Bigg|_{\left(a_1,\ldots,a_n\right) \rightarrow \left(a_n,\ldots,a_1\right)} \ep
\end{equation}
Often the blob has a factorable color structure, i.e.\@
\begin{equation}
  F_{\mu_1\cdots\mu_n}^{a_1\cdots a_n} (k_1,\ldots, k_n) =
    C^{a_1\cdots a_n} F_{\mu_1\cdots\mu_n} (k_1,\ldots, k_n)\ep
\end{equation}
We then can factor out the full color structure from the diagram, implying that the difference between both structures only lies in the color factor in front:
\begin{equation}
  \begin{tikzfigure}{0.6}{0.5}
    \draw[gluon] (0.5,0) -- (0.5,-1.5);
    \draw[gluon] (1,0) -- (1,-1.5);
    \draw[gluon] (2.5,0) -- (2.5,-1.5);
    \node[gluontext] at (1.75,-0.45) {$\ldots$};
    \draw[wilson, -implies] (0,0) -- (3,0);
    \filldraw[wilsontext] (0,0) circle(0.125);
    \filldraw[blob] (1.5,-1.5) circle(1.5 and 0.75);
    \node[blobtext] at (1.5,-1.5) {$F$};
  \end{tikzfigure}
  \!=\; C
  \begin{tikzfigure}{0.6}{0.5}
    \draw[photon] (0.5,0) -- (0.5,-1.5);
    \draw[photon] (1,0) -- (1,-1.5);
    \draw[photon] (2.5,0) -- (2.5,-1.5);
    \node[photontext] at (1.75,-0.45) {$\ldots$};
    \draw[wilson, -implies] (0,0) -- (3,0);
    \filldraw[wilsontext] (0,0) circle(0.125);
    \filldraw[blob] (1.5,-1.5) circle(1.5 and 0.75);
    \node[blobtext] at (1.5,-1.5) {$F$};
  \end{tikzfigure}
  \quad\,\Rightarrow\quad\,
  \begin{tikzfigure}{0.6}{0.5}
    \draw[gluon] (0.5,0) -- (0.5,-1.5);
    \draw[gluon] (1,0) -- (1,-1.5);
    \draw[gluon] (2.5,0) -- (2.5,-1.5);
    \node[gluontext] at (1.75,-0.45) {$\ldots$};
    \draw[wilson, implies-] (0,0) -- (3,0);
    \filldraw[wilsontext] (3,0) circle(0.125);
    \filldraw[blob] (1.5,-1.5) circle(1.5 and 0.75);
    \node[blobtext] at (1.5,-1.5) {$F$};
  \end{tikzfigure}
  \!= \; (-)^n \, \bar{C}
  \begin{tikzfigure}{0.6}{0.5}
    \draw[photon] (0.5,0) -- (0.5,-1.5);
    \draw[photon] (1,0) -- (1,-1.5);
    \draw[photon] (2.5,0) -- (2.5,-1.5);
    \node[photontext] at (1.75,-0.45) {$\ldots$};
    \draw[wilson, -implies] (0,0) -- (3,0);
    \filldraw[wilsontext] (0,0) circle(0.125);
    \filldraw[blob] (1.5,-1.5) circle(1.5 and 0.75);
    \node[blobtext] at (1.5,-1.5) {$F$};
  \end{tikzfigure}
  \ep
\end{equation}
The yellow wavy lines are a reminder that there is no color structure left in the blob.
If the blob isn't color factorable, one can write it as a sum of factorable terms:
\begin{equation*}
  F_{\mu_1\cdots\mu_n}^{a_1\cdots a_n} (k_1,\ldots, k_n) =
    \sum\limits_i C_i^{a_1\cdots a_n} F_{i\, \mu_1\cdots\mu_n} (k_1,\ldots, k_n)\ec
\end{equation*}
such that we can repeat the same procedure as before.

\section{Piecewise Linear Wilson Lines}\label{sec: PiecewiseLinear}
It is not so difficult to generalize the former tricks to piecewise linear Wilson lines. The main difference is that a blob can be connected to several blobs at once. The symmetry properties are then no longer valid for the full blob, but still on a segment-by-segment basis. We can form a basic set of diagrams which span all diagrams involved.
When connecting \eg a $4$-gluon blob, we need to calculate exactly 5 diagrams (independent on the number of segments $M$). These diagrams are:
\begin{equation*}
  \mathcal{U}^J_4, \, \mathcal{U}^J_3\mathcal{U}^K_1, \, \mathcal{U}^J_2\mathcal{U}^K_2, \,
  \mathcal{U}^J_2\mathcal{U}^K_1\mathcal{U}^L_1, \, \text{and }
  \mathcal{U}^J_1\mathcal{U}^K_1\mathcal{U}^L_1\mathcal{U}^O_1.
\end{equation*}
They are the easiest represented schematically:
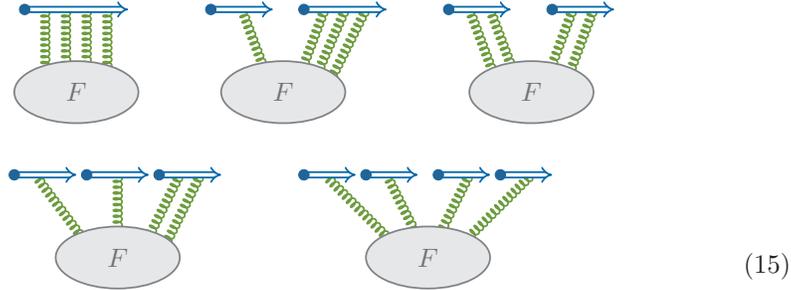
\begin{equation}\label{eq: 4gluondiags}
  \begin{tikzfigure}{0.55}{3.5}
    \draw[gluon] (0.75,0) -- (0.75,-2);
    \draw[gluon] (1.25,0) -- (1.25,-2);
    \draw[gluon] (1.75,0) -- (1.75,-2);
    \draw[gluon] (2.25,0) -- (2.25,-2);
    \draw[wilson, -implies] (0.25,0) -- (2.75,0);
    \filldraw[wilsontext] (0.25,0) circle(0.125);
    \filldraw[blob] (1.5,-2) circle(1.5 and 0.75);
    \node[blobtext] at (1.5,-2) {$F$};
    \begin{scope}[shift={(5,0)}]
      \draw[gluon] (0.5,0) -- (1.25,-2);
      \draw[gluon] (2.5,0) -- (1.75,-2);
      \draw[gluon] (3,0) -- (2.125,-2);
      \draw[gluon] (3.5,0) -- (2.5,-2);
      \draw[wilson, -implies] (-0.25,0) -- (1.25,0);
      \draw[wilson, -implies] (2,0) -- (4,0);
      \filldraw[wilsontext] (-0.25,0) circle(0.125);
      \filldraw[wilsontext] (2,0) circle(0.125);
      \filldraw[blob] (1.5,-2) circle(1.5 and 0.75);
      \node[blobtext] at (1.5,-2) {$F$};
    \end{scope}
    \begin{scope}[shift={(11,0)}]
      \draw[gluon] (0,0) -- (0.75,-2);
      \draw[gluon] (0.5,0) -- (1.25,-2);
      \draw[gluon] (2.5,0) -- (1.75,-2);
      \draw[gluon] (3,0) -- (2.25,-2);
      \draw[wilson, -implies] (-0.5,0) -- (1,0);
      \draw[wilson, -implies] (2,0) -- (3.5,0);
      \filldraw[wilsontext] (-0.5,0) circle(0.125);
      \filldraw[wilsontext] (2,0) circle(0.125);
      \filldraw[blob] (1.5,-2) circle(1.5 and 0.75);
      \node[blobtext] at (1.5,-2) {$F$};
    \end{scope}
    \begin{scope}[shift={(1,-4)}]
      \draw[gluon] (1.5,0) -- (1.5,-2);
      \draw[gluon] (3,0) -- (2,-2);
      \draw[gluon] (3.5,0) -- (2.5,-2);
      \draw[gluon] (-0.5,0) -- (1,-2);
      \draw[wilson, -implies] (0.75,0) -- (2.25,0);
      \draw[wilson, -implies] (-1,0) -- (0.5,0);
      \draw[wilson, -implies] (2.5,0) -- (4,0);
      \filldraw[wilsontext] (-1,0) circle(0.125);
      \filldraw[wilsontext] (2.5,0) circle(0.125);
      \filldraw[wilsontext] (0.75,0) circle(0.125);
      \filldraw[blob] (1.5,-2) circle(1.5 and 0.75);
      \node[blobtext] at (1.5,-2) {$F$};
    \end{scope}
    \begin{scope}[shift={(8.5,-4)}]
      \draw[gluon] (0.5,0) -- (1.5,-2);
      \draw[gluon] (-1,0) -- (1,-2);
      \draw[gluon] (2.5,0) -- (1.5,-2);
      \draw[gluon] (4,0) -- (2,-2);
      \draw[wilson, -implies] (0,0) -- (1.25,0);
      \draw[wilson, -implies] (-1.5,0) -- (-0.25,0);
      \draw[wilson, -implies] (1.75,0) -- (3,0);
      \draw[wilson, -implies] (3.25,0) -- (4.5,0);
      \filldraw[wilsontext] (0,0) circle(0.125);
      \filldraw[wilsontext] (-1.5,0) circle(0.125);
      \filldraw[wilsontext] (1.75,0) circle(0.125);
      \filldraw[wilsontext] (3.25,0) circle(0.125);
      \filldraw[blob] (1.5,-2) circle(1.5 and 0.75);
      \node[blobtext] at (1.5,-2) {$F$};
    \end{scope}
  \end{tikzfigure}
\end{equation}
To relate different path structures, we just use the same trick as before, \eg:
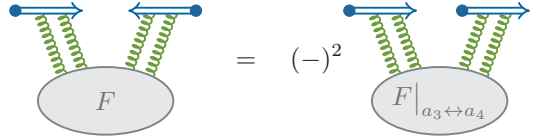
\begin{equation*}
  \begin{tikzfigure}{0.6}{0.75}
    \draw[gluon] (0,0) -- (0.75,-2);
    \draw[gluon] (0.5,0) -- (1.25,-2);
    \draw[gluon] (2.5,0) -- (1.75,-2);
    \draw[gluon] (3,0) -- (2.25,-2);
    \draw[wilson, -implies] (-0.5,0) -- (1,0);
    \draw[wilson, implies-] (2,0) -- (3.5,0);
    \filldraw[wilsontext] (-0.5,0) circle(0.125);
    \filldraw[wilsontext] (3.5,0) circle(0.125);
    \filldraw[blob] (1.5,-2) circle(1.5 and 0.75);
    \node[blobtext] at (1.5,-2) {$F$};
  \end{tikzfigure}
  \quad = \quad
  (-)^2
  \begin{tikzfigure}{0.6}{0.75}
    \draw[gluon] (0,0) -- (0.75,-2);
    \draw[gluon] (0.5,0) -- (1.25,-2);
    \draw[gluon] (2.5,0) -- (1.75,-2);
    \draw[gluon] (3,0) -- (2.25,-2);
    \draw[wilson, -implies] (-0.5,0) -- (1,0);
    \draw[wilson, -implies] (2,0) -- (3.5,0);
    \filldraw[wilsontext] (-0.5,0) circle(0.125);
    \filldraw[wilsontext] (2,0) circle(0.125);
    \filldraw[blob] (1.5,-2) circle(1.5 and 0.75);
    \node[blobtext] at (1.5,-2) {$ F \big|_{a_3\leftrightarrow a_4}$};
  \end{tikzfigure}.
\end{equation*}
To implement this formally, we define a path function $\Phi$ per diagram for a given blob, that gives the color structure in function of the path type. For the leading order 2 gluon blob, this is straightforward:
\begin{subalign}[eq: 2gluonpathconstants]
  \begin{tikzfigure}{0.6}{0.2}
    \draw[gluon] (0.25,0) to[out=-60,in=-120,distance=25] (2,0);
    \draw[wilson,-implies] (-0.25,0) -- (2.25,0);
    \filldraw[wilsontext] (-0.25,0) circle(0.125);
  \end{tikzfigure}
  &:
  \qquad \Phi(J) = C_F,
  \\
  \begin{tikzfigure}{0.6}{0.2}
    \draw[gluon] (0.5,0) to[out=-90,in=-90,distance=30] (2.5,0);
    \draw[wilson,-implies] (-0.5,0) -- (1,0);
    \filldraw[wilsontext] (-0.5,0) circle(0.125);
    \draw[wilson,-implies] (1.5,0) -- (3,0);
    \filldraw[wilsontext] (1.5,0) circle(0.125);
  \end{tikzfigure}
  &:
  \qquad \Phi(J,K) = (-)^{\phi_J + \phi_K}C_F,
\end{subalign}
where $\phi_J$ represents the structure type of the segment:
\begin{equation}
  \phi_J = \begin{cases}
    0 & J = \wilsonup\\
    1 & J = \wilsondown
  \end{cases}\ep
\end{equation}

Let us now introduce a new notation, to indicate a full diagram but without the color content, in which a blob is connected to $m$ Wilson line segments with $n_i$ gluons connected to the $i$-th segment:
\begin{equation}
  \mathcal{W}_{n_m\cdots n_1}^{J_m\cdots J_1}\ep
\end{equation}
We can now give a symbolic expression for a 4-gluon blob connected to a piecewise linear Wilson line:\cite{VanderVeken:2014kna,VanderVeken:2014eda}
\begin{multline}\label{eq: 4gluonblobresult}
  \mathcal{U}^4 = 
    \sum_J^M \Phi_4\mathcal{W}_{4}^{J} +
    \sum_{J=2}^M \sum_{K=1}^{J-1}
      \left[ \Phi_{3\,1} \mathcal{W}_{3\,1}^{JK} + \Phi_{2\,2} \mathcal{W}_{2\,2}^{JK} \right]
    +
    \sum_{J=3}^M \sum_{K=2}^{J-1} \sum_{L=1}^{K-1} \Phi_{2\,1\,1} \mathcal{W}_{2\,1\,1}^{JKL}\\
    +
    \sum_{J=4}^M \sum_{K=3}^{J-1} \sum_{L=2}^{K-1} \sum_{O=1}^{L-1} 
      \Phi_{1\,1\,1\,1} \mathcal{W}_{1\,1\,1\,1}^{JKLO} +
    \text{symm.}
\end{multline}
Both the $\Phi_{n_i\cdots}$ and $\mathcal{W}_{n_i \cdots}$ can be calculated independently of the path structure, giving a result depending on $n_J$, $r_J$ and $\phi_J$, which then can easily be ported to different path structures.

\section{Example Calculation}
Let us illustrate our framework with a small example, viz.\@ the LO 2-gluon blob. At any order, there are 2 possible 2-gluon diagrams:
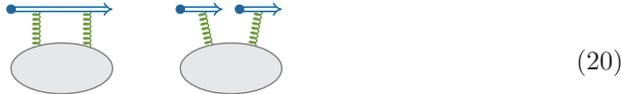
\begin{equation}
\begin{tikzfigure}{0.45}{0.8}
  \draw[gluon] (0.75,0) -- (0.75,-1.75);
  \draw[gluon] (2.25,0) -- (2.25,-1.75);
  \filldraw[blob] (1.5,-1.75) circle(1.5 and 0.75);
  \draw[wilson,-implies] (0,0) -- (3,0);
  \filldraw[wilsontext] (0,0) circle(0.125);
  \begin{scope}[shift={(5,0)}]
    \draw[gluon] (0.625,0) -- (1,-1.75);
    \draw[gluon] (2.325,0) -- (2,-1.75);
    \filldraw[blob] (1.5,-1.75) circle(1.5 and 0.75);
    \draw[wilson,-implies] (0,0) -- (1.25,0);
    \filldraw[wilsontext] (0,0) circle(0.125);
    \draw[wilson,-implies] (1.75,0) -- (3,0);
    \filldraw[wilsontext] (1.75,0) circle(0.125);
  \end{scope}
\end{tikzfigure}
\end{equation}
At NLO the blob is just an LO gluon propagator.
The color factors are given in Eq.\ \eqref{eq: 2gluonpathconstants}. We show here the result for the second diagram at $r_J=r_K$. If both segments are on-LC, the result is:
\begin{equation}
	\mathcal{W}_{1\,1}^{JK} \big|_{\text{LC}} =
		\frac{\alpha_s}{2\pi}
		\left( \frac{1}{\epsilon^2} + \frac{\pi^2}{3} \right)
		\left(
			\frac{n_K\!\cdot\! n_J}{2}\frac{\mu^2}{\eta^2}
		\right)^{\!\epsilon}
		\ep
\end{equation}
If both segments are off-LC, the result can be expressed in function of the angle $\chi$:
\begin{equation}
	\cosh \chi = \frac{n_K\cdot n_J}{|n_K|\,|n_J|}\ep
\end{equation}
For simplicity we normalize the directions, i.e.\@ $|n_K|=|n_J|$. This gives:
\begin{subalign}
\mathcal{W}_{1\,1}^{JK} \big|_{\text{\sout{LC}}} &=
	\frac{\alpha_s}{2\pi}\chi\text{coth}\chi\,
	\left(\frac{1}{\epsilon}+\Upsilon\right)
	\left[
		\frac{1}{4} n_K^2n_J^2 \sinh^2\!\chi\,
		\frac{\mu^2}{\eta^2}
	\right]^{\!\epsilon} \ec\\
\Upsilon &= 2\ln 2 \!+\! \ln n_K^2\!+\!2\ln(1\!+\!\exp{\chi})\!+\!\chi
		\!-\!\frac{1}{\chi}\left(
			\text{Li}_2\,\exp{\chi}\!-\!\text{Li}_2\,\exp{-\chi}
		\right) \ep
\end{subalign}
If $|n_K|\neq|n_J|$, only $\Upsilon$ is affected and will be a bit more involving.

\section*{Acknowledgements}
I am very grateful to the organisers of the SPIN 2014 conference for creating such a fruitful environment. Furthermore I would like to thank I.O.~Cherednikov, M.~Echevarria, L.~Gamberg, A.~Idilbi, T.~Mertens, A.~Prokudin and P.~Taels for useful discussions and insights.

\bibliographystyle{ws-ijmpcs}
\bibliography{Bibliography}

\end{document}